\documentclass[12pt]{iopart}
\usepackage{graphicx}
\begin{document}
\title[Mobile Phone Faraday Cage]{Mobile Phone Faraday Cage}
\author{M M J French$^1$}
\address{$^1$ Physics Department, Clifton College, 32 College Road, Clifton, Bristol. BS8 3EZ}
\ead{mail@matthewfrench.net}
\begin{abstract}
A Faraday cage is an interesting physics phenomena where an
electromagnetic wave can be excluded from a volume of space by
enclosure with an electrically conducting material. The practical
application of this in the classroom is to block the signal to a
mobile phone by enclosing it in a metal can! The background of the
physics behind this is described in some detail followed by a
explanation of some demonstrations and experiments which I have
used.
\end{abstract}
Submitted to: \emph{Phys. Ed.} \maketitle
\section{Static Electromagnetic Fields}
A Faraday cage is a hollow enclosure of conducting material (eg. a
metal) which blocks out static electric fields. Imagine a static
electric field applied to a hollow metal cylinder. The metal
cylinder contains charges which are free to move around within the
metal. As long as the electric field is applied the charges will
move around until they get to the surface of the metal but can then
move no further. The charges will collect at the surface and produce
an electric field within the metal cylinder which exactly opposes
the applied field. This mechanism can be compared with a simple
electrical circuit with a 1.5V cell: one side of the cell is at a
higher potential than the other so charges move to cancel this out.
Thus a static electric field can be excluded from a volume of space
by surrounding the space with a conducting material. However, a
mobile phone signal is not a static: it is an electromagnetic wave
so we now look at the effect of a sinusoidal electric field.
\section{Sinusoidal Electromagnetic Fields}
In the case of a sinusoidal electric field the case is a little more
complex. Imagine a simple sinusoidal electromagnetic field whose
behaviour is described by Maxwell's equations:
\begin{eqnarray}
\nabla \cdot \mathbf{E} &=& 0 \\
\nabla \cdot \mathbf{B} &=& 0 \\
\nabla \times \mathbf{E} &=& -\frac{\partial \mathbf{B}}{\partial t} \\
\nabla \times \mathbf{B} &=& \frac{1}{c^2}\frac{\partial
\mathbf{E}}{\partial t}+\mu_0\sigma \mathbf{E}
\end{eqnarray}
where $\mathbf{E}$ is the electric field, $\mathbf{B}$ is the
magnetic field, $t$ is time, $c$ is the speed of light, $\mu_0$ is
the permeability of free space and $\sigma$ is the electrical
conductivity of the metal making up the enclosure. Consider a
special type of wave which is an unbounded plane wave. By a plane
wave we mean that over planes perpendicular to the direction of the
propagation of the wave, all field quantities (such as amplitude,
frequency etc.) are constant. By unbounded we mean that the planes
are infinite. If the direction of propagation of the wave is taken
to be the $x$ direction, then the wavefronts (or the planes
mentioned earlier) are parallel to the $yz$ plane. This means that
the partial differential of $\mathbf{E}$ or $\mathbf{B}$ with
respect to $y$ or $z$ is zero:
\begin{eqnarray}
\frac{\partial}{\partial y}&=&0 \\
\frac{\partial}{\partial z}&=&0
\end{eqnarray}
This reduces Maxwell's equations to:
\begin{eqnarray}
\frac{\partial E_z}{\partial x} &=& \frac{\partial B_y}{\partial t} \\
\frac{\partial E_y}{\partial x} &=& -\frac{\partial B_z}{\partial t} \\
\frac{\partial B_z}{\partial x} &=& -\frac{1}{c^2}\frac{\partial E_y}{\partial t} \\
\frac{\partial B_y}{\partial x} &=& \frac{1}{c^2}\frac{\partial
E_z}{\partial t}-\mu_0\sigma E_y
\end{eqnarray}
Further, if we choose that the $\mathbf{E}$ field is along the $y$
axis then $E_z=0$. This implies that the $\mathbf{B}$ field is along
the $z$ axis and
\begin{eqnarray}
\frac{\partial B_y}{\partial x}&=&0 \\
\frac{\partial B_y}{\partial t}&=&0
\end{eqnarray}
Finally, Maxwell's equation further reduce to:
\begin{eqnarray}
\frac{\partial E_y}{\partial x}&=& -\frac{\partial B_z}{\partial t}
\\
\frac{\partial B_z}{\partial x}&=&-\frac{1}{c^2}\frac{\partial
E_y}{\partial t}-\mu_0\sigma E_y
\end{eqnarray}
$B_z$ (or $E_y$) can be eliminated from the previous two equations
to give the wave equation (which is the same in each case):
\begin{equation}
\frac{\partial^2 E_y} {\partial x^2}=\frac{1}{c^2}\frac{\partial ^2
E_y}{\partial t^2}+\mu_0\sigma\frac{\partial E_y}{\partial t}
\end{equation}
Making this relevant to a mobile phone signal we now focus on
monochromatic waves (i.e. waves with a single frequency) with the
form:
\begin{equation}\label{monochromaticequation}
E_y(x,t)=E_{y0}e^{i(\mathbf{k}x-\omega t)}
\end{equation}
where $i=\sqrt{-1}$, $\mathbf{k}$ is the wave number and $\omega$ is
the angular frequency of the wave. Substituting this into the wave
equation above gives us:
\begin{equation}
\mathbf{k}^2=\frac{\omega^2}{c^2}+i\omega \mu_0\sigma
\end{equation}
In a metallic conductor the conduction current will dominate over
that caused by the displacement of charges due to the field, so that
\begin{equation}
\mathbf{k}^2\approx i\omega \mu_0\sigma
\end{equation}
Using the identity that
\begin{equation}
i^{\frac{1}{2}}=\frac{1+i}{\sqrt{2}}
\end{equation}
this simplifies to
\begin{equation}
\mathbf{k}=\frac{1+i}{\delta}
\end{equation}
where
\begin{equation}
\delta=\left(\frac{2}{\mu_0\sigma\omega}\right)^\frac{1}{2}
\end{equation}
where $\delta$ is the penetration (or skin) depth. Substituting this
back into the monochromatic wave (equation
\ref{monochromaticequation}) above gives
\begin{equation}
E_y(x,t)=E_{y0}e^{-\frac{x}{\delta}}e^{i(\frac{x}{\delta}-\omega t)}
\end{equation}
This indicates that the wave attenuates as it travels through a
metal in the $x$ direction with a penetration depth of $\delta$. For
steel with conductivity of $10^6$ Siemens per meter and at the
frequency of a mobile phone signal (around 1GHz) this gives a
penetration depth of the order 10-100$\mu$m. The theory in this
section was adapted from \cite{Duffin}.
\par
It is important to realise that the penetration depth is not the
thickness of surrounding material which reduces the electric field
to a magnitude of zero, rather it is the characteristic distance
over which the magnitude decreases exponentially. An exponential
decrease means that the rate of decrease is proportional to the
magnitude of the signal with constant of proportionality $1/\delta$:
\begin{equation}
\frac{dE_y(x)}{dx} = -\frac{1}{\delta}E_y(x)
\end{equation}
where $E_y(x)$ is the maximum wave amplitude as a function of $x$,
the distance into the metal.
\par
\begin{figure}\begin{center}
\includegraphics[width=10cm]{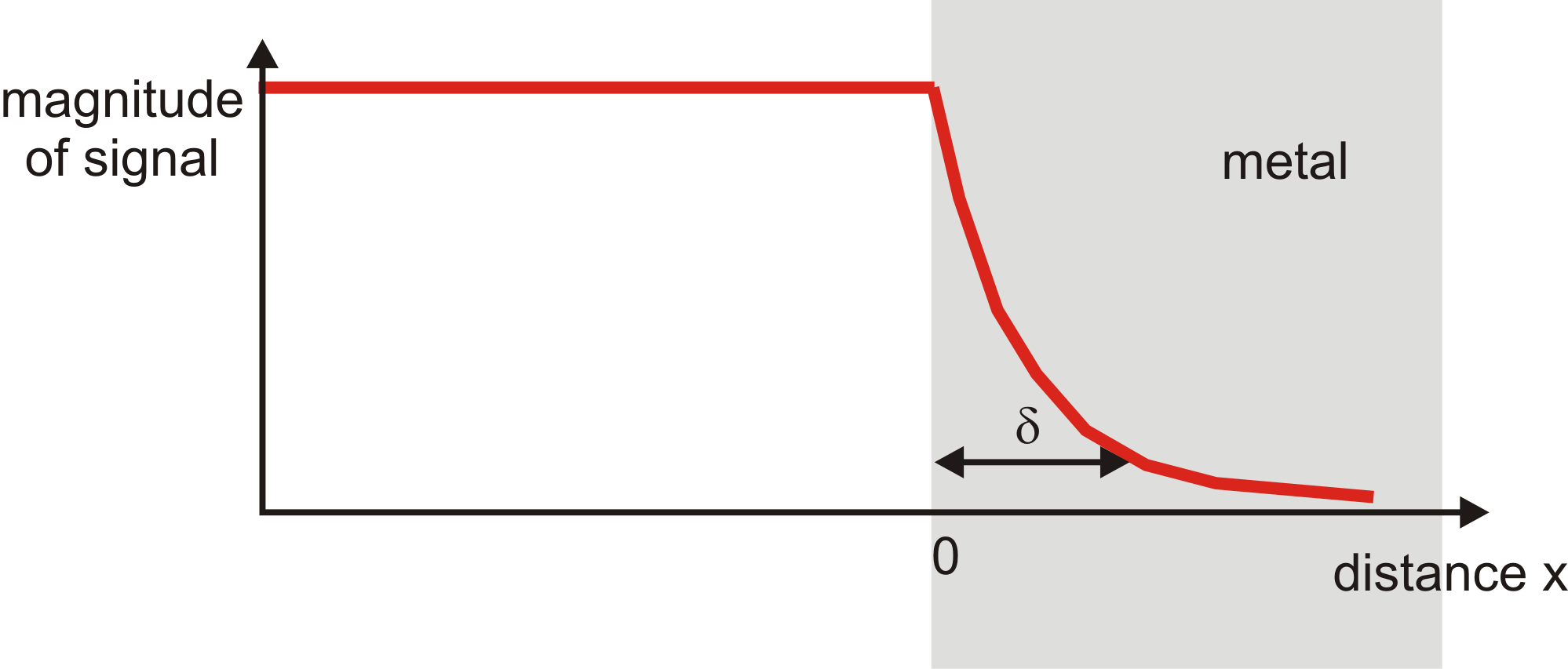}
\caption {The exponential decrease of the magnitude of an electric
field inside a conductor} \label{exponential}
\end{center}\end{figure}

At a distance of $x=0$ into the metal
\begin{equation}
e^{-x/\delta}=e^{-0/\delta}=1
\end{equation}
so the amplitude of the wave is not yet decreased. The thickness at
which the magnitude will drop to 10\% of its original value is given
by:
\begin{eqnarray}
e^{-x/\delta}&=&0.1 \\
-\frac{x}{\delta}&=&\ln{0.1}\\
x&=&2.3\delta
\end{eqnarray}
By a similar calculation a drop to a 1\% will occur after a distance
of 4.6$\delta$ into a metal and to 0.1\% after 6.9$\delta$. See
figure \ref{exponential}.
\par
It is not easy to ascertain what electric field magnitude (signal
strength) is required for a certain mobile phone to connect to a
network - it will depend on the phone. But if we guestimate that it
will be between 1\% and 0.1\% of the original signal magnitude we
need a thickness of up to 0.7mm of steel.
\par
In a certain location, the magnitude of the electric field from the
local mobile phone mast will depend on environmental and
geographical factors as well as which mobile network the phone is
connected to.
\section{Developing an Experiment}
A first attempt at demonstrating the Faraday cage using a mobile
phone was performed using two empty soup tins pressed together with
the phone placed inside. The enclosed phone was dialed from another
phone to see if it rang. This method worked fairly reliably with any
mobile phone and at any location. I have used this in lessons to
demonstrate the behaviour of electromagnetic waves for GCSE pupils
(age 14\,-\,16 years) and A level pupils (age 16\,-\,18 years). An
interesting side point is that a metal foil lined bag is sometimes
used by shoplifters to prevent security tags on goods setting off
alarms as they a smuggled out of shops.
\begin{figure}\begin{center}
\includegraphics[width=10cm]{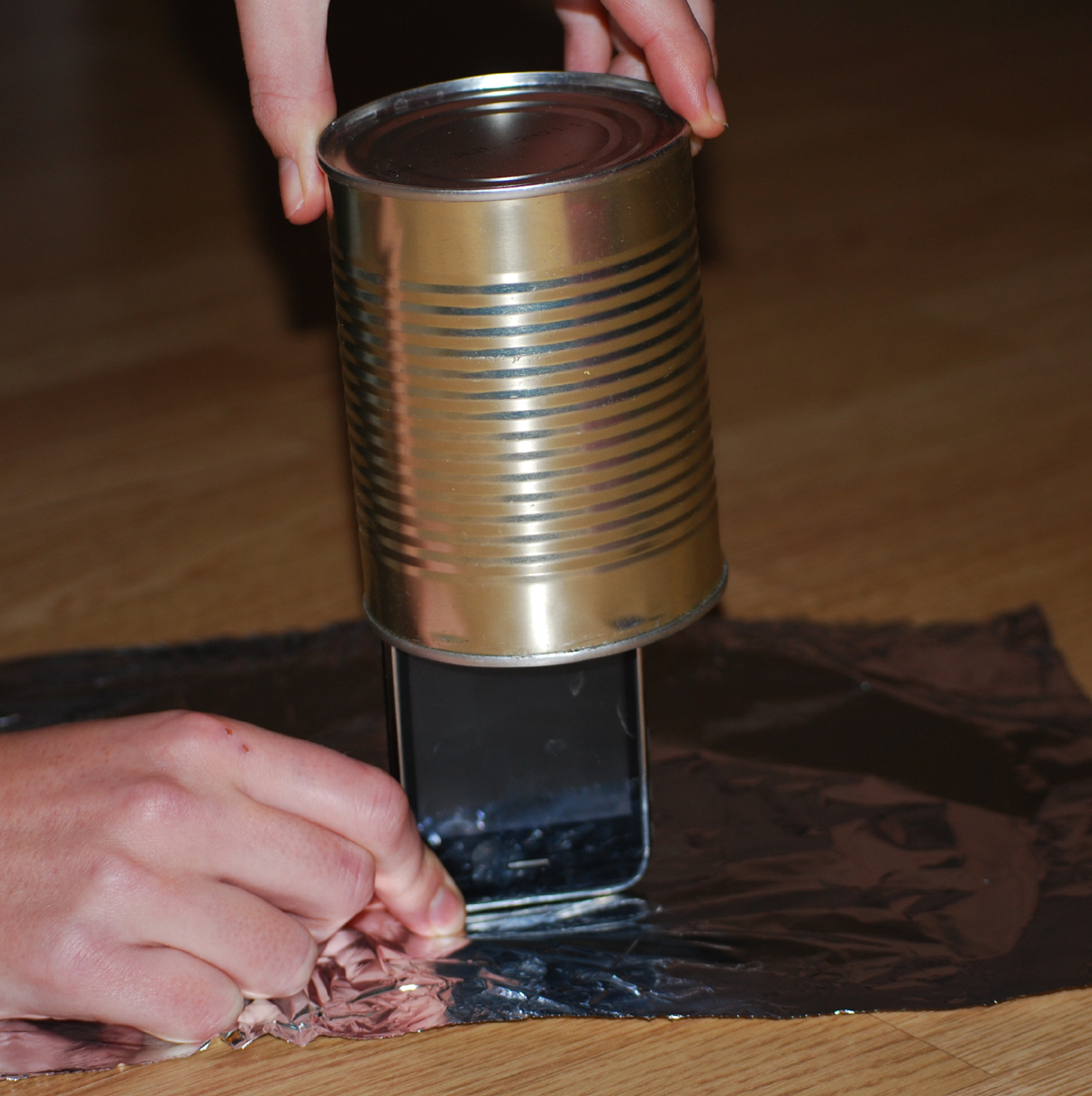}
\caption {Placing a phone in the tin.} \label{photo2}
\end{center}\end{figure}
\begin{figure}\begin{center}
\includegraphics[width=10cm]{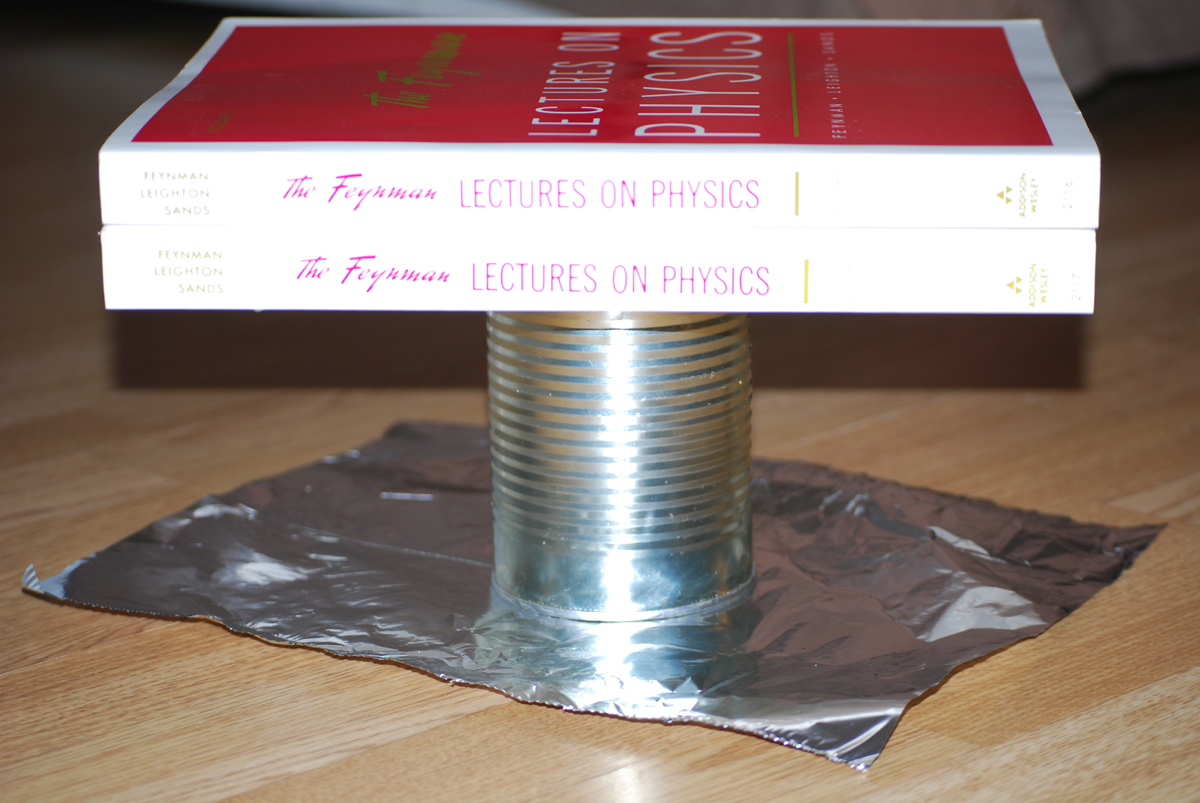}
\caption {Phone in the tin, with a textbook to ensure there are no
unwanted gaps in the Faraday cage.} \label{photo1}
\end{center}\end{figure}
\par
For a mobile phone signal the wavelength and frequencies are given
in Table 1 for the networks which operate their own base stations in
the UK \cite{sitefinder}.

\begin{table}
\begin{center}
\begin{tabular}{ccc}
  \hline
  Network & Frequency & Wavelength \\
\hline
  Orange & 1800MHz & 17cm \\
  O$_2$ & 900MHz & 33cm \\
  Vodafone & 900MHz & 33cm \\
  T-Mobile & 1800MHz & 17cm \\
  \hline
\end{tabular}
\caption{Table of mobile phone networks, frequencies and wavelengths
in the UK}
\end{center}
\end{table}
Note that the 2100\,MHz signals listed in \cite{sitefinder} for a
number of networks is the frequency over which data is sent and not
voice calls. The pupils can either be given the above table of
frequencies and wavelengths, or if their ability is high enough,
they can be asked to find the wavelength for their phone, given the
frequency.
\par
Interestingly the Faraday cage still works if there are gaps in the
conductor, providing their maximum diameter is `significantly
smaller' than the wavelength of the electromagnetic waves. I have
developed this into an experiment in which the pupils determine what
`significantly smaller', in this case, means. Working in small
groups of 2-3, a single tin (opened at one end) is used. When the
phone is placed in the tin the open end is covered with metal foil,
see figure \ref{photo2}. Sometimes the foil is too thin to totally
block the signal (as discussed above) for one or more networks, but
this depends on the relative location of the classroom and the
mobile phone mast. This `failure' can be turned into a learning
opportunity but explaining the situation to the group of pupils and
guiding them to use a phone from a different network. A good `seal'
is required between the tin, which is upturned on the desk, and the
foil which lies on the desk. A text book or two placed on top of the
tin usually suffices, see figure \ref{photo1}.
\par
One pupil from the group uses their phone to call the phone in the
tin/foil enclosure. Once the pupils have got the phone to fail to
ring, they are ready for the next step. A small hole needs to be
placed in the foil and the phone tested again to see if it will
ring. The hole size then needs to be gradually increased until the
phone in the tin starts ringing. The largest `diameter' of the hole
is measured each time and is recorded in a table by the pupils along
with an indication of whether the phone rang or not.
\par
The final part of the lesson gives the pupils the opportunity to
contribute their results to a large table on the board and to
discuss and think about their results. The board table has columns
for phone network, wavelength, and maximum hole size that stops the
phone ringing.
\par
If there are around 8-10 groups in the class, I have found that I
usually get sufficiently good results to draw at least a qualitative
conclusion: that the size of the hole can be bigger for the networks
with the longer wavelength of radiation. Thus, a mobile phone on a
network with a short wavelength (Orange or T-Mobile) will be best if
you are tying to use the phone in a lift!
\par
As a further angle of investigation, the location, frequency and
network of every phone mast in the UK can be found online
\cite{sitefinder}. This information may also be available in other
countries. In some cases where the mast of a certain network is
based on or very close to the school site the examples of Faraday
cages described in this article are not effective as the signal
strength is too strong. That is, the magnitude of the signal is not
sufficiently reduced by the thickness of metal in a piece of foil or
a metal can.

\section{Acknowledgements}
I am grateful for the encouragement I received from Michael Hibbert
at The Ellen Wilkinson School for Girls, Acton, London in developing
and using this experiment and for the opportunity to demonstrate
this at Clifton College in late 2009. Miriam Segelman was invaluable
in helping with some of the photographs contained in this work as
well as the initial idea for the application to a lift.

\begin{center}
\includegraphics[width=4cm]{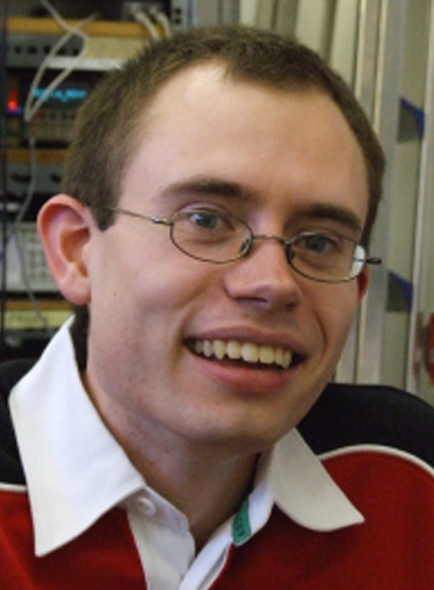}
\end{center}
\par
After completing a PhD at the University of Bristol
and a PGCE at Imperial College, Matthew French has recently started
teaching physics at Clifton College, Bristol.
\section{References}
\bibliographystyle{unsrt}
\bibliography{references}
\end{document}